\documentclass[fleqn,12pt,twoside]{article}
\usepackage{graphicx}

\newcommand{\AmS}{{\protect\the\textfont2
  A\kern-.1667em\lower.5ex\hbox{M}\kern-.125emS}}

\title{Spinodal in asymmetric nuclear matter}

\author{J\'er\^ome Margueron and Philippe Chomaz\\
GANIL CEA/DSM - CNRS/IN2P3 BP 5027 F-14076 Caen CEDEX 5, France}

\begin{document}

\maketitle

\begin{abstract}
The phase diagram of nuclear matter is quite rich - it shows such 
phenomena as phase-transitions, formation of condensates, clustering, etc.
From the analysis of the spinodal instability, one can learn about the region of 
liquid-gas coexistence  in nuclear matter at low densities 
and finite isospin asymmetries. In a recent paper, we have shown
that asymmetric nuclear matter at sub-nuclear densities should
undergo only one type of instability.
The associated order parameter is dominated by the isoscalar density and so
the transition is of liquid-gas type. The instability goes in the direction
of a restoration of the isospin symmetry leading to a fractionation
phenomenon. 
\end{abstract}


The nuclear interaction is very similar to the Van der Waals potential which
acts between molecules. For this reason,
below saturation density, the nuclear interaction is also expected to
lead to a liquid-gas phase transition \cite{ber80}. Recently, a converging
ensemble of experimental signals seems to have established the phase
transition. One is the spinodal decomposition \cite{bor01} which consider
volume instabilities (domain of negative incompressibility).
One expect that the system which enter in such a forbiden region will 
favorably breakup into nearly equal-sized ``primitive'' fragments in relation
to the wavelengths of the most unstable modes \cite{ayi95}.
How this simple picture is modified by the asymmetry charge ? Can we expect
new signals related to the collision of very asymmetric nuclei ?

\section{Stability analysis}

Let us consider asymmetric nuclear matter (ANM) 
characterized by a proton and a neutron densities $\rho
_{i}=$ $\rho _{p}$, $\rho _{n}$. These densities can be transformed in a set
of 2 mutually commuting charges $\rho _{i}=$ $\rho _{1}$, $\rho _{3}$ where $%
\rho _{1}$ is the density of baryons, $\rho _{1}=\rho _{n}+\rho _{p},$ and $%
\rho _{3}$ the asymmetry density $\rho _{3}=\rho _{n}-\rho _{p}$. In
infinite matter, the extensivity of the free energy implies that it can be
reduced to a free energy density~: $F(T,V,N_{i})=V\mathcal{F}(T,\rho _{i}).$
The system is stable against separation into two phases if the free energy
of a single phase is lower than the free energy in all two-phases
configurations. This stability criterium implies that the free energy
density is a convex function of the densities $\rho _{i}$. A local necessary
condition is the positivity of the curvature matrix~: 
\begin{equation}
\left[ \mathcal{F}_{ij}\right] =\left[ \frac{\partial ^{2}\mathcal{F}}{%
\partial \rho _{i}\partial \rho _{j}}|_{T}\right] \equiv \left[ \frac{%
\partial \mu _{i}}{\partial \rho _{j}}|_{T}\right]  \label{eq5}
\end{equation}
where we have introduced the chemical potentials $\mu _{j}\equiv \frac{%
\partial F}{\partial N_{j}}|_{T,V,N_{i}}=\frac{\partial \mathcal{F}}{%
\partial \rho _{j}}|_{T,\rho _{i\not{=}j}}$.

\begin{figure}[p]
\center
\includegraphics[scale=0.7]{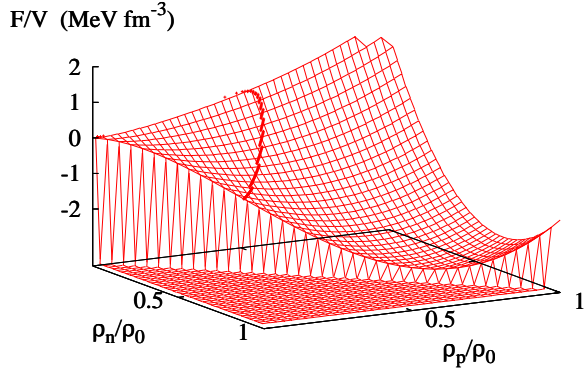}
\caption{This figure represents the energy surface as a function of the
densities $\rho_n$ and $\rho_p$ for the SLy230a interaction. The contour
delimitate the spinodal area. }
\label{fig1}
\end{figure}

We represent in Fig.~\ref{fig1} the energy surface as a function of $\rho
_{n}$ and $\rho _{p}$, deduced from SLy230a Skyrme interaction~\cite{cha97}.
In the symmetric case ($\rho _{n}=\rho _{p}$), one can see the negative
curvature of the energy which defines the spinodal area, whereas in pure
neutron matter ($\rho _{p}=0$), no negative curvature and so no spinodal
instability are predicted. We can also notice that the isovector density
dependence is almost parabolic illustrating the positivity of $\mathcal{F}%
_{33}$.

\begin{figure}[p]
\center
\includegraphics[scale=0.4]{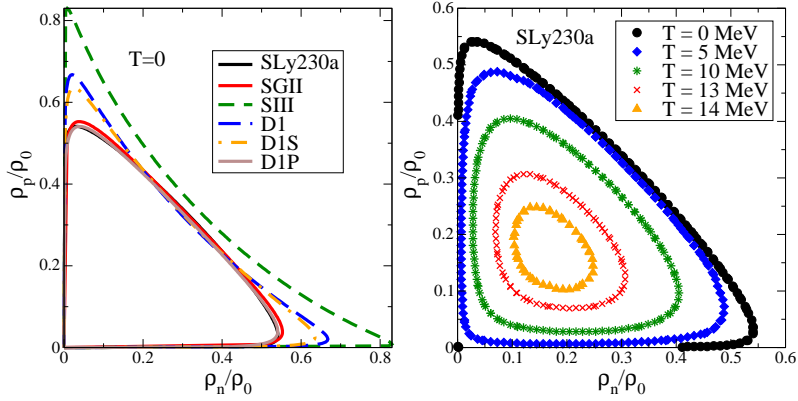}
\caption{This two figures are a projection of the spinodal contour in the
density plane : left, for Skyrme (SLy230a~\protect\cite{cha97}, SGII~ 
\protect\cite{ngu81}, SIII~\protect\cite{bei75}) and Gogny models (D1~%
\protect\cite{gog75}, D1S~ \protect\cite{ber91}, D1P~\protect\cite{far99}) ;
right, temperature dependence of the spinodal zone computed for the SLy230a
case.}
\label{fig2}
\end{figure}

We show in Fig.~\ref{fig2} several aspects of the spinodal contour defined
as the region where the matrix $\left[ \mathcal{F}_{ij}\right]$ is negative.
In the left part is plotted the spinodal contour in ANM for
several forces. It exhibits important differences. In the case of SLy230a
force (as well as SGII, D1P), the total density at which spinodal
instability appears decreases when the asymmetry increases whereas for SIII
(as well as D1, D1S) it increases up to large asymmetry and finally
decreases. We observe that all forces which fulfill the global requirement
that they reproduce symmetric nuclear matter (SNM) equation of state as well
as the pure neutron matter calculations, leads to the same curvature of the
spinodal region. We can appreciate the reduction of the instability when we
go away from SNM. However, large asymmetries are needed to induce a sizable
effect. The temperature dependence of the spinodal contour can be
appreciated in the right part of Fig.~\ref{fig2}. As the temperature
increases the spinodal region shrinks up to the critical temperature for
which it is reduced to SNM critical point. However, up to a rather high
temperature ( $5$ $\mathrm{MeV}$) the spinodal zone remains almost identical
to the zero temperature one.

\begin{figure}[tbph]
\center
\includegraphics[scale=0.4]{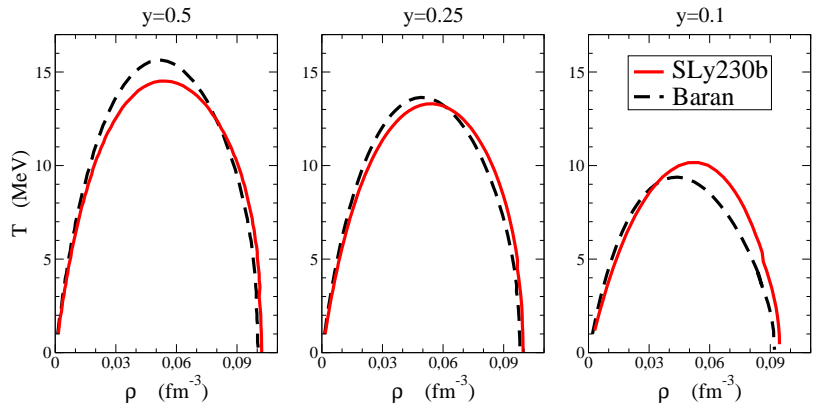}
\caption{Comparison between the results obtained by Skyrme simplified Baran et al
interaction~\cite{bar01} and the one obtained with SLy230b Skyrme interaction.}
\label{fig7}
\end{figure}

Almost all theoretical predictions has been made with simplified Skyrme
interactions : in medium nucleonic masses are taken as the free masses and
spin exchange terms proportionnal to $x_i$
are not explicitly treated. In our case, we have included the
standard terms of the interactions that we refer to. Fig.~\ref{fig7} shows a
comparison between our calculation and one of those simplified interactions 
(used by Baran et al~\cite{bar01}). 
In SNM (y=0.5), exploring high temperatures means exploring the k dependance of
the single particle potential, hence the k-effective mass of the nucleons in
the medium. On the counter part, increasing the asymmetry means being sensitive
to the isospin dependance of the effective mass. As the k-effective mass used by
Baran et al is independant of the asymmetry parameter but is only a function of
the total density, we can conclude that the effective mass (according to SLy230b)
reduces the critical temperature in SNM by about 1 MeV while it increases the 
critical temperature by 1 MeV in very asymmetric matter. This
comparison shows that for qualitative predictions, a simple interaction
is enough but quantitative predictions requires the standard interaction.

\section{Analysis of the curvature matrix $[\mathcal{F}_{ij}]$}

In the considered two-fluids system, the $[\mathcal{F}_{ij}]$ is a $2*2$
symmetric matrix, so it has 2 real eigenvalues $\lambda ^{\pm}$~\cite{bar01}%
~: 
\begin{equation}
\lambda ^{\pm}=\frac{1}{2}\left( \mathrm{Tr}\left[ \mathcal{F}_{ij}\right]
\pm \sqrt{\mathrm{Tr}\left[ \mathcal{F}_{ij}\right] ^{2}-4\mathrm{Det}\left[ 
\mathcal{F}_{ij}\right] }\right)  \label{eq23}
\end{equation}
associated to eigenvectors $\mathbf{\delta \rho }^{\pm}$ defined by ($i\neq
j $) 
\begin{equation}
\frac{{\delta \rho }_{j}^{\pm}}{{\delta \rho }_{i}^{\pm}}=\frac{\mathcal{F}%
_{ij}}{\lambda ^{\pm}-\mathcal{F}_{jj}}=\frac{\lambda ^{\pm}-\mathcal{F}_{ii}%
}{\mathcal{F}_{ij}}  \label{eq25}
\end{equation}
Eigenvectors associated with negative eigenvalue indicate the direction of
the instability. It defines a local order parameter since it is the
direction along which the phase separation occurs. The eigen values $\lambda 
$ define sound velocities, $c$, by ${c}^{2}=\frac{1}{18m}\rho _{1}\,\lambda
. $ In the spinodal area, the eigen value $\lambda $ is negative, so the
sound velocity, $c$, is purely imaginary and the instability time $\tau $ is
given by $\tau =d/|c|$ where $d$ is a typical size of the density
fluctuation.

The requirement that the local curvature is positive is equivalent to the
requirement that both the trace ($\mathrm{Tr}[\mathcal{F}_{ij}]=\lambda
^{+}+\lambda ^{-})$ and the determinant ($\mathrm{Det}[\mathcal{F}%
_{ij}]=\lambda ^{+}\lambda ^{-})$ are positive 
\begin{equation}
\mathrm{Tr}[\mathcal{F}_{ij}]\geq 0,\hbox{ and }\mathrm{Det}[\mathcal{F}%
_{ij}]\geq 0  \label{eq6}
\end{equation}
The use of the trace and the determinant which are two basis-independent
characteristics of the curvature matrix clearly stresses the fact that the
stability analysis should be independent of the arbitrary choice of the
thermodynamical quantities used to label the state e.g. $(\rho _{p}$, $\rho
_{n})$ or $(\rho _{1}$, $\rho _{3})$.
If Eq.~\ref{eq6} is
violated the system is in the unstable region of a phase transition. Two
cases are then possible~: i) only one eigenvalue is negative and one order
parameter is sufficient to describe the transition or ii) both eigenvalues
are negative and two independent order parameters should be considered
meaning that more than two phases can coexist.
For ANM below saturation density, the case ii) never occurs since the
asymmetry energy has always positive curvature ($\mathcal{F}_{33}$). 
A complete discussion is presented in~\cite{mar03}.

\begin{figure}[htbp]
\center
\includegraphics[scale=0.3]{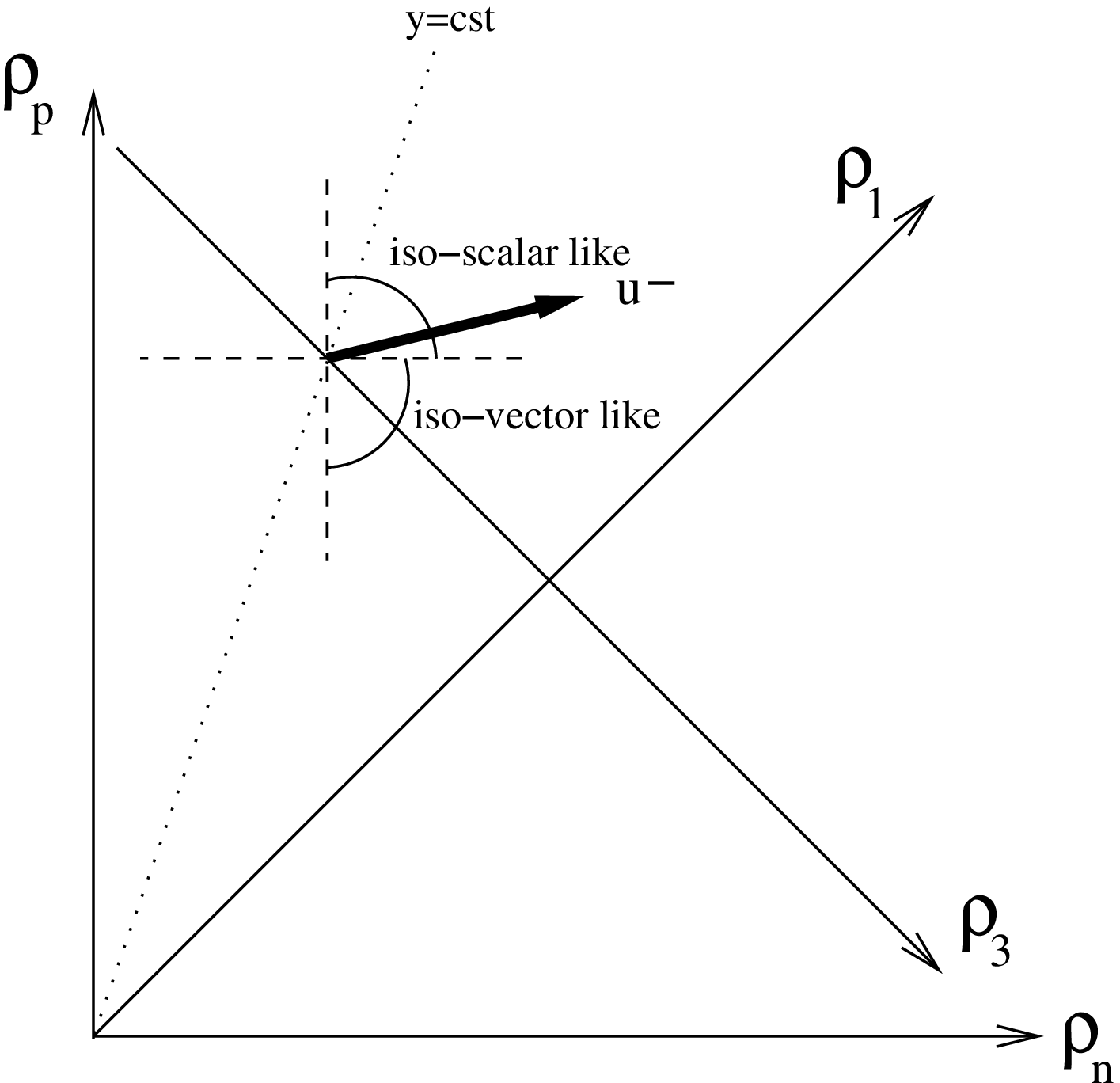}
\includegraphics[scale=0.4]{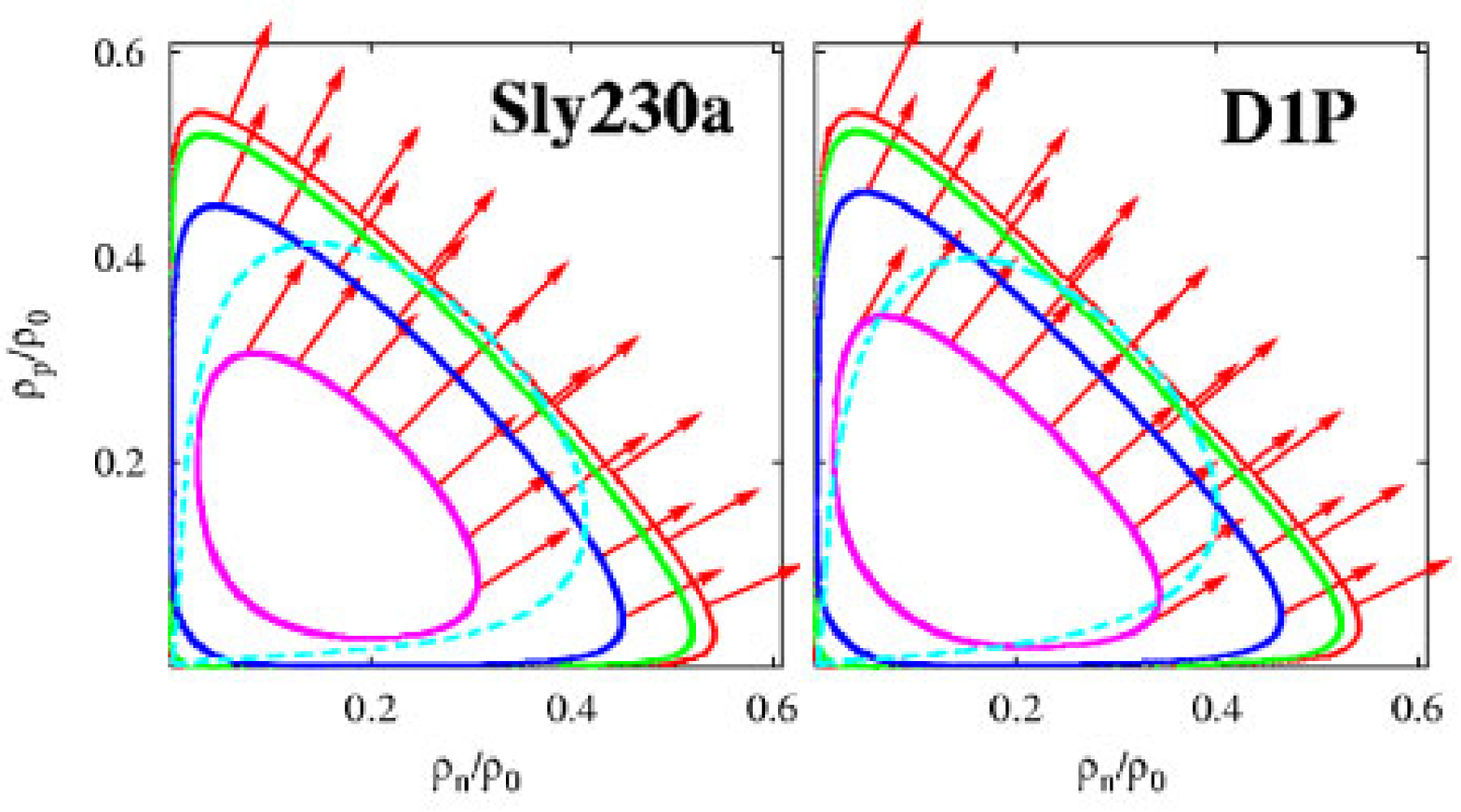}
\caption{On the left, we illustrate the generalization of the definition of
isoscalar and isovector modes.
Various contour of equal imaginary sound velocity are also represented 
for SLy230b and D1P interactions. The more internal curve
correspond to the sound velocity $i0.09c$, after comes $i0.06c$, $i0.03c$
and finally 0, the spinodal boarder. We observe that in almost all the spinodal
region the sound velocity is larger than $0.06c$.
The arrows indicate the direction of
instability. The mechanical instability is also indicated (dotted line).}
\label{fig3}
\end{figure}

\smallskip Let us now focus on the direction of the instability. If $\mathbf{%
\delta \rho }^{-}$ is along $y$=cst then the instability does not change the
proton fraction. For symmetry reasons pure isoscalar $(\delta \rho _{3}=0)$
and isovector $(\delta \rho _{1}=0)$ modes appears only for SNM so it is
interesting to introduce a generalization of isoscalar-like and
isovector-like modes by considering if the protons and neutrons move in
phase ($\delta \rho _{n}^{-}\delta \rho _{p}^{-}>0$) or out of phase ($%
\delta \rho _{n}^{-}\delta \rho _{p}^{-}<0$). 
We propose a generalization of the definition of
isoscalar and isovector modes in ANM. According to the left graph of 
Fig.~\ref{fig3}, the
instability is of isoscalar type if its direction points in the direction of
the first bissectix with an absolute value angle less that 45 degrees, while
it is of isovector kind if its directions points in the direction of the
second bissectrix with an absolute value angle less that 45 degrees.
The two figures on the rigth part of Fig.~\ref{fig3} shows the
direction of instabilities along the spinodal boarder and some
iso-instability lines. We observed that instability is always almost along
the $\rho _{1}$ axis meaning that it is dominated by total density
fluctuations even for large asymmetries. 
This shows that the unstable direction is of isoscalar nature as expected 
from the attractive interaction between proton-neutron~\cite{bar01,mar03}. 
The total density is therefore the dominant
contribution to the order parameter showing that the transition is between
two phases having different densities (i.e. liquid-gas phase transition).
The angle with the $\rho _{1}$ axis is almost constant along a constant $y$
line. This means that as the matter enters in the spinodal zone and then
dives into it, there are no dramatic change in the instability direction
which remains essentially a density fluctuation. Moreover, the unstable
eigenvector drives the dense phase (i.e. the liquid) towards a more
symmetric point in the density plane. By particle conservation, the gas
phase will be more asymmetric leading to the fractionation phenomenon. Those
results are in agreement with recent calculation for ANM~\cite{bar01} and
nuclei~\cite{col02}.

\section{Are they new signals in asymmetric matter~?}

A frequent discussion can be found in the literature~\cite{mul95,bao97}.
It is argued that asymmetric nuclear matter do not only present a mechanical
instability for which the density is the order parameter but also a broader
chemical instability associated with fluctuations of the matter isospin
content  \cite{mul95}.  Indeed, it is usually argued that it exists a
region in which the compressibility at constant isospin asymmetry
is negative (see Fig.1 of \cite{cho02}) leading to the interpretation
that the system is mechanically unstable. Above a maximum asymmetry the
isotherms at constant asymmetry does not presents any back bending leading
to the idea that the system is mechanically stable. However, looking at the
equilibrium of the chemical potentials one can see that above this maximum
asymmetry for mechanical instabilities the system may amplify fluctuations
in the proton neutron concentration leading to a second instability region
usually called chemical instabilities.      

However, we have recently shown that this splitting of the spinodal region
into two types of instabilities, a mechanical and a chemical one, is not
correct and that ANM present only one type of instability \cite{mar03}
(hereafter called the first argument).
This result is robust because it can be related to the density dependance
of the asymmetry energy reproduced by several models of the nuclear 
interaction (Skyrme, Gogny, Brueckner).
In a recent proceeding \cite{cho02}, we have discussed that figures like 
Fig.1 of \cite{cho02} may lead to the unphysical separation of two area in the
spinodal.
Indeed, these two regions are artefact in the sense that it comes from
a 2-dimensionnal projection of the 3-dimensionnal representation of 
the free energy (see for instance Fig.\ref{fig1}). At the entrance of
the spinodal region, only one direction is unstable (the direction pointed
out by the eigen vector, see Fig.\ref{fig3}). The further the system sink
inside the spinodal, the wider become the unstable directions and each
point in the spinodal is a saddle point.
Deep inside the spinodal, the particuliar direction $y=\rho_p/\rho=$const
becomes also unstable but the system do not enter inside a new phase as it is
often claimed.

Here, we will give a third argument against the artifical separation of
chemical and mechanical instabilities. This argument is based on the
relations first demonstrated by Baran et al~\cite{bar01}. It rely the
eigen modes to the chemical and mechanical definitions. It reads:
\begin{eqnarray}
\frac{\partial P}{\partial \rho _{1}}|_{T,y} &=&\frac{\lambda ^{-}}{\sqrt{t}}%
(t\cos \beta +\sin \beta )^{2}+\frac{\lambda ^{+}}{\sqrt{t}}(t\sin \beta
-\cos \beta )^{2} \label{eqq1}\\
\frac{\partial \mu _{p}}{\partial y}|_{T,P} &=&\rho _{n}\lambda ^{+}\lambda
^{-}\left( \frac{\partial P}{\partial \rho _{1}}|_{T,y}\right) ^{-1} \label{eqq2}
\end{eqnarray}
where $\beta =1/2\arctan \mathcal{F}_{np}/(\mathcal{F}_{pp}-\mathcal{F}_{nn})$
and $t=\rho _{n}N_{0}^{n}/\rho _{p}N_{0}^{p}$. 
In SNM, $t=1$ and $\beta=\pi/4$. Then, the relation between the eigen values 
$\lambda^+$, $\lambda^-$ and the definition of mechanical and chemical instabilities
is trivial: below saturation density, SNM is unstable toward density fluctuations 
which means that $\lambda^-$ or $\frac{\partial P}{\partial \rho _{1}}|_{T,y}$
become negative. Beyond saturation density, some interactions manifest an
instability in the isospin channel, and once again, it can be associated to
the negativity of $\lambda^+$ or $\frac{\partial \mu _{p}}{\partial y}|_{T,P}$.
Hence, it is totally equivalent to discuss the negativity og the eigen values
$\lambda^+$, $\lambda^-$ or the onset of a chemical and mechanical instability.
But, in ANM, the simplicity of this equivalence is not preserved as it is shown by
Eq.~\ref{eqq1} and Eq.~\ref{eqq2}.
Hence, one should trust only the eigen analysis of the curvature matrix $[%
\mathcal{F}_{ij}]$ and
this analysis have shown that only one of the two eigen values ($\lambda^-$)
changes its sign.
Then, there is one unstable mode in ANM below saturation density.

\section{Conclusion}

Finally, we have presented three arguments in favor of the fact that
ANM does not present two types of spinodal
instabilities, a mechanical and chemical, but only one which is dominantly
of isoscalar nature. This means
that the instability is always dominated by density fluctuations and so can
be interpreted as a liquid-gas separation. The instabilities tend to restore
the isospin symmetry for the dense phase (liquid) leading to the
fractionation of ANM. We have shown that changing the asymmetry up to $\rho
_{p}<3\rho _{n}$ does not change quantitatively the density at which
instability appears, neither the imaginary sound velocity compared to those
obtained in SNM. The quantitative predictions
concerning the shape of the spinodal zone as well as the instability times
depends upon the chosen interaction but converge for the various forces
already constrained to reproduce the pure neutron matter calculation.

\end{document}